\documentclass[twocolumn,showpacs,pre,aps,superscriptaddress]{revtex4}
\usepackage[utf8]{inputenc}
\usepackage[english]{babel} 
\usepackage{graphicx} 
\usepackage{listings} 
\usepackage{amssymb} 
\usepackage{amsmath} 
\usepackage{longtable}
\usepackage{multirow} 
\usepackage{booktabs}
\usepackage{pst-all} 
\usepackage{pstricks-add}
\usepackage{units} 
\usepackage{listings} 
\usepackage{subfigure}

\usepackage{xcolor}

\begin{document} 

\title{Initial State Independent Equilibration {at the Breakdown} of the Eigenstate Thermalization
Hypothesis}

\author{Abdellah Khodja}

\email{akhodja@uos.de}

\affiliation{Fachbereich Physik, Universit\"at Osnabr\"uck,
             Barbarastrasse 7, D-49069 Osnabr\"uck, Germany}

 \author{Daniel Schmidtke}
\email{danischm@uos.de}

\affiliation{Fachbereich Physik, Universit\"at Osnabr\"uck,
             Barbarastrasse 7, D-49069 Osnabr\"uck, Germany}

\affiliation{Fachbereich Physik, Universit\"at Osnabr\"uck,
             Barbarastrasse 7, D-49069 Osnabr\"uck, Germany}

\author{Jochen Gemmer}

\email{jgemmer@uos.de}

\affiliation{Fachbereich Physik, Universit\"at Osnabr\"uck,
             Barbarastrasse 7, D-49069 Osnabr\"uck, Germany}

 \begin{abstract}
 This work aims at understanding the interplay between the Eigenstate Thermalization Hypothesis (ETH), initial state
independent equilibration and 
 quantum chaos in systems that do not have a direct classical counterpart. It is based on numerical investigations of
asymmetric Heisenberg spin ladders with 
 varied interaction strengths between the legs, i.e., along the rungs.
 The relaxation of the  energy difference between the legs is investigated. Two different parameters, both intended to
quantify the degree of 
 accordance with the ETH, are computed. {Both indicate violation of the ETH  at
large interaction strengths but at
different thresholds.} Indeed the energy 
 difference
 is found not to relax independently of its initial value above some critical interaction strength which coincides with
one of the thresholds. At the same point
 the level statistics shift from Poisson-type to Wigner-type. Hence the system may be considered to become integrable
again in 
 the strong interaction limit.
\end{abstract}


\maketitle

\section{introduction}
\label{intro}

Even in closed quantum systems one may observe relaxation, in spite of time evolution being generated by
unitary operations. This statement has many aspects, it may imply that a reduced density matrix of a part of some system
approaches a thermal state \cite{goldstein, 
Popescu, lychkovskiy2010, Gogolin}, it may imply that expectation values of specific observables evolve more or less against constant values
\cite{neumann, Reimann}, or it may additionally even 
imply that these constant
values do not depend on the initial state  \cite{neumann, Zelditch, Shnirelman}. It is this latter initial state
independence (ISI) which is in the focus of the 
paper at hand. 
Among the concepts addressing this issue is the so called  eigenstate
thermalization hypothesis (ETH) \cite{deutsch1991,srednicki1994}. According to the ETH, the
expectation values of a typical observable $\hat{D}$ as computed from energy eigenstates $|n\rangle$ should be
a smooth function of energy  $E_n$ 
, i.e., $\langle n | \hat{D} | n \rangle \approx
\langle n' | \hat{D} | n' \rangle= \bar{D}_{eq}(E)$ if $E_n \approx E_{n'}\approx E$. 
As is well known, if the ETH applies, the long time averages of all expectation value $ \text{Tr} \{\hat{\rho} \,
\hat{D}(t)\}$ corresponding to any initial state $\hat{\rho}$ that lives 
inside some energy shell, are equal \cite{rigol2008}. Furthermore, if the spectrum of the Hamiltonian $\hat{H}$ is
``non-resonant'' (roughly
speaking: any energy difference 
occurs only once)
and the $\rho_{nn}$ are fairly spread over the energy shell, then the actual $ \text{Tr} \{\hat{\rho} \,
\hat{D}(t)\}$ deviates from
its long time average very rarely
 \cite{reimann2008}. Hence, the three features, i. ETH agreement, ii. non-resonant spectrum and iii. dilute eigenstate
occupation, guarantee the initial state
 independent (ISI) relaxation of the expectation value towards a specific ``equilibrium'' value. Since this behavior is
observed and expected for practically 
 all physical relaxation phenomena, the question arises whether the above three features apply to all those
situations. This question has been approached 
 from the perspective of quantum chaos, and various rigorous results exist that suggest i. and ii. apply to systems
which have a direct classical counterpart
 which is chaotic \cite{Zelditch}. Much less is known for systems that do not have a direct classical counterpart
\cite{srednicki1994}. However, while the three 
 features are 
 sufficient for ISI, they are not necessary in a mathematical sense: classes of initial states exist that exhibit ISI
even though the ETH may not apply. Some
 papers put much more emphasis on the extremely high relative frequency with which ISI may be expected if initial state
are drawn essentially at random from 
 some prescribed sets, rather than 
 on the ETH \cite{linden, wehner, Reimann, lychkovskiy2010}. However, it may be the case that the relative frequency of
initial states from the above sets, 
 that exhibit a significant deviation of the respective 
 expectation value from its equilibrium value at all, even at $t=0$, is very low. In this case results of
relative-frequency-type would imply the existence
 of a majority of states the expectation values of which start
 and remain in equilibrium. However, no conclusions on state, that actually do exhibit non-equilibrium expectation
values in the beginning, could be drawn. 
 
 In view of this, a class of initial states that are specifically tailored to exhibit largely deviating expectation
values, but live in 
 narrow energy shells at the same time has recently been suggested \cite{abdell}. These states (which will be explained
in detail below) have been termed
 microcanonical observable displaced (MOD) states.  MOD states are ``natural'' in the sense that they may be viewed as
results of state determination
 according to Jayne's principle under the conditions of a given expectation value and the state living in an certain
energy shell. For the remainder of this 
 paper we focus on dynamics as resulting from such initial MOD states.
 
 In order to set the content of the paper at hand in context to the state of research, we need to specify our notion of
ETH in somewhat more detail. Consider
 \begin{equation}
\bar{D} = \sum_{n=1}^d p_n \, D_{nn} \, , \quad \Sigma^2 =
\sum_{n=1}^d p_n \, D_{nn}^2 - \bar{D}^2 \, , \label{ETH1}
\end{equation}
where $D_{nn}$ are diagonal matrix elements with respect to\ the Hamiltonian
eigenstates $| n \rangle$ with eigenvalues $E_n$, $p_n \propto
e^{-(E_n - \bar{E})^2 / 2 \sigma^2_E}$ is a probability distribution
centered at $\bar{E}$, and $d$ is the Hilbert-space dimension. The
quantities $\bar{D}(\bar{E},\sigma_E)$ and $\Sigma(\bar{E},\sigma_E)$
are obviously functions of $\bar{E}$ and an energy width $\sigma_E$.
Routinely, the ETH is said to be fulfilled if $\Sigma$ is small for 
small $\sigma_E$. 
{Technically, ISI equilibration is only guaranteed for all initial states from a given energy window
$\sigma_E$ at $\bar{E}$ iff
$\Sigma = 0$, \cite{Rigol1}.  If instead a finte-size scaling scheme is employed, i.e., $\Sigma \rightarrow 0$ for $N
\rightarrow \infty$ ($N$ being the ``system-size'', e.g., the 
number of spins, etc.) not only the Hamiltonian, but also the observable $\hat{D}$ often scales with $N$ in some fashion
as well (cf. below). In this context it is 
relevant to note that ISI equilibration is only ensured if  $\Sigma$ itself approaches zero in the  limit of large
system sizes, the vanishing of, e.g.,  
$\Sigma^2/N$, is not sufficient.
However, in our analysis we do not intend to demonstrate ISI equilibration for all initial states of a given energy
window but rather focus on  the afore mentioned special class of states, i.e., MOD states. Below we will employ a 
quantifier which is closely
related to $\Sigma$ but independent of the scaling of the observable itself.}

{In order to put our work into perspective, we list some  exemplary results from the  literature on the
generic scaling properties of $\Sigma$, 
and classify our results in relation to  these. In the context of translational 
invariant, solid-state type observables and models, the observable is frequently  defined to scale ``extensively'', 
like, e.g., a current, a total kinetic energy, etc.
In this case, there is substantial numerical evidence that  $\Sigma$ scales more or less as $\Sigma \propto
d_{eff}^{-\gamma}$, where $d_{eff}$ is the effective dimension, i.e., the number of states within the respective energy
shell
and $\gamma$ is some constant 
\cite{masud, Jacobs-Vishnapasi, steinigeweg}. Note, however, that the scaling behavior of $\Sigma$  generally depends
strongly on whether the observable
itself is defined to scale extensively, intensively  or else. 
Since the
effective dimension $d_{eff}$ usually increases (exponentially) with system size, 
$\Sigma$ tends to zero with increasing 
system size whenever
$\gamma >0$. The existing literature contains numerous examples exhibiting  $0 \leq \gamma \leq 1/2$ for few body
observables.
$\gamma=0$ appears to be related to integrable systems, $\gamma=1/2$ for fully chaotic systems, where both definitions
may vary depending on the
respective
authors \cite{shastry}. 
}

{In  Ref. \cite{steinigeweg}, however, also examples are discussed with $\gamma < 0$ for extensive
observables, i.e.,
the unscaled $\Sigma$ itself increases with system size. This occurs for  spin and energy currents in the  (integrable)
Heisenberg chain at some parameter regime.
In contrast to that, in Ref. \cite{abdell}  the same Heisenberg
chain model is investigated while the considered observable is different, namely the  energy difference between two
parts of the chain. The latter observable is also extensive. 
Nonetheless, in this case $\Sigma=\text{const}$, i.e., $\gamma=0$ is found. As already mentioned above, for $\gamma \leq
 0$ ISI equilibration is not guaranteed, 
i.e., the construction of initial states for which the expectation value of  $\hat{D}$ does not decay towards the
corresponding equilibrium ensemble value is 
always possible. Thus we will  primarily focus on the question whether such a ``incomplete decaying'' occurs for
MOD states for models and observables exhibiting 
 $\gamma \leq  0$.}

{For the remainder of this paper we follow \cite{abdell} in  investigating the (extensive) energy
difference between certain system sections. We present a model that 
allows for a continuous tuning of the scaling parameter $\gamma$ all the way from $\gamma >0$ to $\gamma < 0$, i.e.,
depending on some system parameter, $\Sigma$ 
either decreases, remains constant or increases under upscaling of the system-size.
We also  follow \cite{abdell} in considering an alternative parameter $v$ for the ``prediction'' of 
ISI for MOD states:
\begin{align}
v=\left(\frac{\displaystyle\Sigma}{\displaystyle\delta_D}\right)^2 
\end{align}
 $\delta_D^2$ denotes the spectral variance of the observable $\hat{D}$ in the
respective energy shell, i.e., $\delta^2_D=\Sigma^d_{n=1} p_n (\hat{D}^2)_{nn}-(\bar{D })^2$ ({the bars
over $D$ symbolize the averaging as defined
in Eq. (\ref{ETH1}))}. Note that  $v$ is, other than $\Sigma$, dimensionless. Thus for the scaling behavior of  $v$, the
scaling behavior of the observable itself
(extensive, intensive or else) is irrelevant. 
It has been proposed that ISI for MOD states occurs, if and only if $v \rightarrow 0 $ in
the limit of large systems. 
Although closely related to the ETH, the $v$ criterion is not quite the same, for it may possibly approach zero even if 
$\Sigma$ increases,  if $\delta_D^2$ increases more quickly.} 

The present paper is organized as follow: In section \ref{model}, we shortly introduce the investigated models and the
addressed observable, i.e., 
the energy difference. In section \ref{ethinvest}, we present the
results on $\Sigma, v$ in dependence of the tuning parameter $\kappa$ and show the existence of two different
regimes based on the scaling behavior of the former.  The computations in Sec. \ref{isi} clarify whether
MOD states give rise to ISI relaxation of the energy difference and how this relates to the results of
Sec.\ref{ethinvest}.  We discuss the integrability of the model at hand in Sec. \ref{integ} using the generalized Brody
parameter, whose behavior turns to be correlated to the ETH parameter $v$. Finally,  we close with summary and
conclusion.

\section{Models and observables}
\label{model}

\begin{figure}[htbp]

\includegraphics[width=0.07\textwidth]{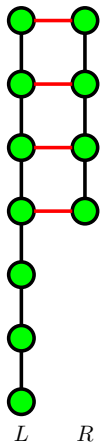}
\caption{Schematic visualization of the investigated spin-ladder. To suppress symmetry, the number of spin in the
right chain is taken to be different than the left chain in the following systematic fashion $N_L=2N_R-1$. Note that
both chains only interact by vertically opposing sites (here indicated by red rungs). }
\label{fig1}

\end{figure}

The model we address in this work is an asymmetrical Heisenberg-ladder which consists of two XXZ spin-$1/2$
chains of different length, coupled along some ``rungs'' with interaction strength $\kappa$, cf. Fig. \ref{fig1}. The
total Hamilton operator is given as follow
\begin{align}
\label{H}
\hat{H}_{}=\hat{H}_{L}+\hat{H}_{R}+\kappa\hat{H}_{I} 
\end{align}
where
\begin{align}
\label{H1}
\hat{H}_{L,R}
=\sum_{i=1}^{N_L,N_R}[(\hat{S}_i^x\hat{S}_{i+1}^x+\hat{S}_i^y\hat{S}_{i+1}^y)+\Delta\hat{S}_i^z\hat{S}_{i+1}^z]
\end{align}
describes the two side-chains. $\Delta$ denotes the anisotropy parameter, which is kept at $\Delta=0.1$ throughout the
entire investigation. $N_L,N_R$ are the numbers of spins with
respect to left and right spin chain respectively, thus the total number of spins is $N=N_L+N_R$. We choose $N_L=2N_R-1$
in
order to suppress symmetry, since for the symmetrical case the ETH
is trivially fulfilled. The two subsystems are allowed to exchange energy through the interaction Hamiltonian, which
reads 
\begin{align}
\label{H2}
\hat{H}_{I}
=\sum_{i=1}^{N_L}[(\hat{S}_i^{x,L}\hat{S}_{i}^{x,R}+\hat{S}_i^{y.L}\hat{S}_{i}^{y,R})+\Delta\hat{S}_i^{z,L}\hat{S}_{i}^{
z,R}] \nonumber
\end{align}
Obviously, at any non-zero  $\kappa$, this model is not accessible by a Bethe ansatz. Thus, in this sense it is always
non-integrable. For a more detailed discussion of integrability, see Sec. \ref{integ}

The observable we are going to investigate is the energy difference operator $\hat{D}
= \hat{H}_L - \hat{H}_R $. Our motivation for this particular choice stems form the intuitive example of two the
cups
of coffee brought into contact with each other, where one anticipate more or less an uniform energy density throughout
the system. The  model (\ref{H}) has been numerically studied in detail for weak interactions, where the ETH
turned out to be
valid \cite{abdell},

\section{Computation of ETH-quantifying parameters}
\label{ethinvest}

Numerical computation of ETH related data like $\Sigma, \bar{D}$, etc.  for large systems is not a trivial task.
Usually, it requires exact diagonalization \cite{steinigeweg,beugeling2014,abdell} that
is limited to rather small system sizes. Thus, we apply a recently
suggested method \cite{steinigeweg2014-1} that is based on dynamical
typicality and allows for the extraction of information about ETH
from the temporal propagation of pure states. This propagation can be
performed by iterative algorithms such as Runge-Kutta \cite{elsayed2013,
steinigeweg2014-1, steinigeweg2014-2}, Chebyshev \cite{deraedt2006,
deraedt2007} etc.\ and is feasible for larger system sizes. We
use a Chebyshev iterator with reasonably small
time step in order to improve the computation accuracy. Due to typicality-related reasons, the so-computed
quantities $\bar{D}$ and $\Sigma$ are subjected to statistical errors.
These errors, however, turn out to be small. Apart from the model size $N=14$, which is treated by exact
diagonalization 
(LAPACK-routine), all data in this section 
has been computed using the above technique. This way we are able to address systems up to $N=26$.

\begin{figure}[htbp]
\includegraphics[width=0.45\textwidth]{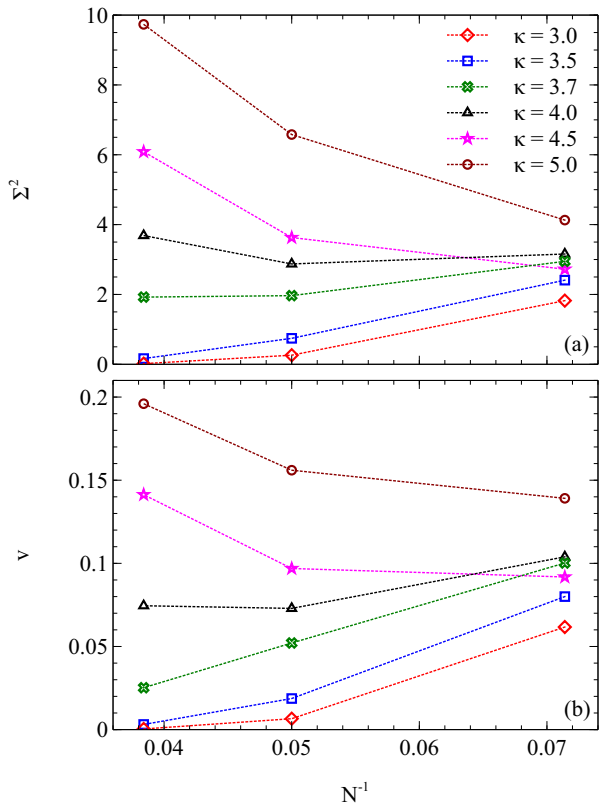}
\caption{{Comparison of the finite size scalings of the  "bare" ETH-fluctuation $\Sigma$ (a) and the
scaled ETH parameter $v$ (b) in dependence of
the interaction strength $\kappa$. Both quantifiers vanish in the large system limit for small  $\kappa$ while they reach constant values (or possibly even diverge)
for large  $\kappa$. For  $\Sigma$ the transition between the  two different regimes appears to occur at  $\kappa \approx 3.7$. Whereas  
for  $v$ the transition appears to occur at  $\kappa \approx 4.0$.}}
\label{fig7}
\end{figure}

We focus on a narrow energy shell around $E=0$ which is the energy regime with the highest density of states. More
precisely we choose
$\bar{E}=0$ and $\sigma_E = 0.6$. To set this into perspective, the total energy range of this type of model is on the
order of $N$.
The results for the ETH-fluctuations $\Sigma$ are depicted in figure \ref{fig7}(a). 

{Figure  \ref{fig7}(a) indicates that there are two clearly distinct regimes: Above  $\kappa = 3.7$ a
convergence of the ETH parameter
$\Sigma$ to zero appears unlikely, even though the presented data may not allow for the precise determination of
$\Sigma$ in the large system limit.
At $\kappa = 3.7$ a simple linear scaling also indicates a non-zero $\Sigma$ in the large system limit, even though the
presented data may no be 
entirely conclusive.
However, below $\kappa = 3.5$ Fig.  \ref{fig7}(a) clearly indicates 
$\Sigma \rightarrow 0 $ for $N \rightarrow \infty$.}
Thus for {$\kappa \ge 3.7$} the considered model potentially shows no ISI
relaxation of the energy differences between the
two chains. This is to 
be contrasted with the overall concept of heat conduction or {the second law of thermodynamics}, which
demands that energy
eventually distributes evenly over all parts of a system, 
regardless of how uneven it was distributed in the beginning. 

Figure  \ref{fig7}(b) displays the ETH parameter $v$ which has been suggested as a ``detector'' of ISI
relaxation \cite{abdell}. 
While the behavior of  $\Sigma$ and $v$ may appear similar at first sight, there are relevant differences: for all 
$\kappa \le 3.7$ $v$ clearly 
vanishes in the limit of large systems, whereas at $\kappa = 4.0 $ a more or less constant scaling occurs
{for $N \geq 20$. Thus the ``switching''
from vanishing to non-vanishing values in the large system limit appears to occur at $\kappa \approx  3.7$ for 
$\Sigma$, whereas it occurs at $\kappa \approx  4.0$ 
for $v$. }


\section{investigation of isi-relaxation of energy differences from initial mod-states}
\label{isi}

In this section we are going to discuss the long time behavior of $\langle D(t) \rangle$ as resulting from initial
states
from the  class
of MOD states. The latter are defined as follows:
\begin{equation}
\label{mod}
 \rho_\text{MOD}:\propto e^{(-(\hat{H} - H_0\hat{1})^2 -\beta^2 (\hat{D}-D_0\hat{1})^2)/ 2 \sigma^2}
\, .
\end{equation}
These states may be viewed as being based on Jayne's
principle: They  represents the maximum-entropy states under given means
and variances for the energy  and the observable. Since $\hat{H}$ and $\hat{D}$ do not commute with each other, it is
not possible
to generate  states with 
arbitrary values of total energy 
energy $\langle \hat{H} \rangle_{MOD}$, energy difference  $\langle \hat{D} \rangle_{MOD}$ and the respective variances.
However, by
tuning the
parameters 
$H_0, D_0, \beta, \sigma$, carefully, we are able to prepare initial states
having a well-defined energy width $\sigma_E \approx 0.6$ around $E=0$ and exhibiting initial
expectation values for the  observable ${D}(0)=\langle \hat{D} \rangle_{MOD}$, which deviate strongly from their
corresponding 
equilibrium values $D_{eq}$; more quantitatively,  $\langle D \rangle_{MOD}$ reaches about $50\%$ of the
difference between its highest possible value and its long-time
average. To enable such strong deviations we choose $D_0 = \pm (N_L - 2)$ throughout all
investigations. This is to be contrasted with various quench scenarios \cite{rossini,rigol2012}, where the initial
deviations
from the equilibrium value are rather small.  Note also that such 
MOD states do not necessarily feature a smooth probability
distribution with respect to the corresponding energy eigenbasis, as required by Ref.\ \cite{ikeda2013}
in order to establish ISI. 

Neither constructing states of the MOD-type (\ref{mod}) nor propagating such states according to the Schroedinger
equation
is numerically simple 
for larger systems. Thus, we instead prepare and evolve a corresponding pure state $| \phi_\text{MOD} \rangle$, which
exhibits, up to very small 
statistical errors the same  $\langle D(t) \rangle$ as $\rho_\text{MOD}$:
\begin{equation}
| \phi_\text{MOD} \rangle = \langle \varphi | \rho_\text{MOD} |
\varphi \rangle^{-1/2} \, \rho_\text{MOD}^{1/2} \, | \varphi \rangle
\, , \label{init}
\end{equation}
where $|\varphi \rangle$ is a random state drawn according to the
unitary invariant (Haar-) measure. This concept relies on dynamical typicality and has been explained and applied in 
e.g, \cite{christian, elsayed2013, abdell, steinigeweg}.

\begin{figure}[h!]
   \centering
   \includegraphics[width=0.42\textwidth]{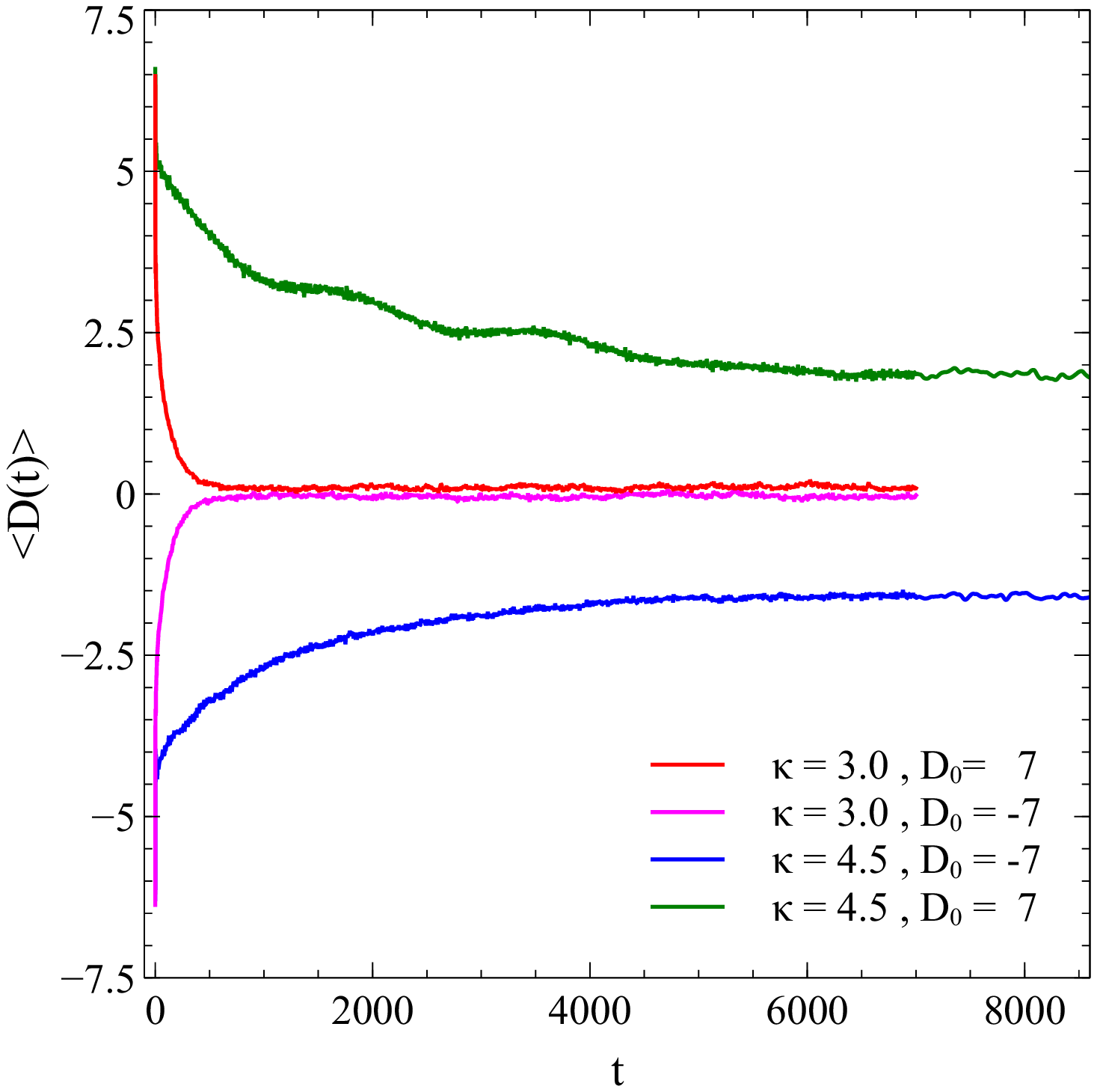}
    \caption{Time evolution of $\langle D(t) \rangle_{MOD}$ for various interaction strengths starting at $D_0=\pm 7$
for
systems of size $N=26$. In case of $\kappa = 3.0$, ISI equilibration is obvious but for $\kappa = 4.5$ the expectation
value converges against finite offsets indicating that some portion of the initial value persists (see text for
details). Note that due to slightly different $|D(0)|$ occurring for $\kappa = 4.5$ the "equilibrium
values" for corresponding curves also slightly differ.}
    \label{evo}
\end{figure}

\begin{figure}[h!]
   \centering
   \includegraphics[width=0.42\textwidth]{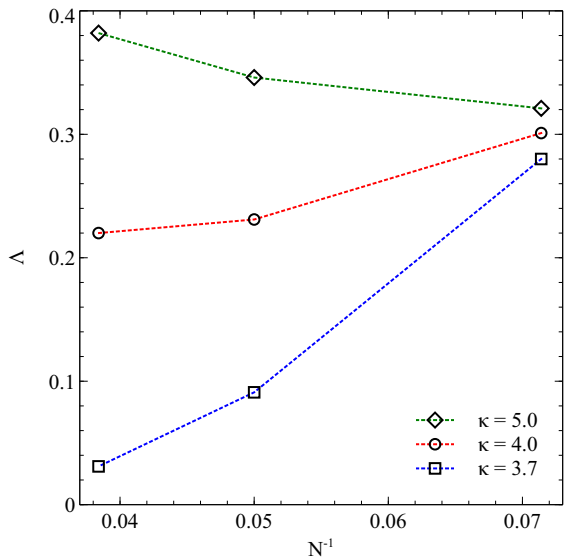}
    \caption{Scaling of $\Lambda$ with various interaction strengths and system sizes. At  $\kappa = 3.7$
the
ISI-quantifying $\Lambda$ tends to zero indicating ISI equilibration whereas for both larger interaction strengths
$\Lambda$ remains finite in the limit of large systems quantifying deviations from the equilibrium value $D_{eq}$. This
behavior corresponds rather accurately to the finite size scaling behavior of $v$, cf. Fig. \ref{fig1}. Statistical errors are of order of symbol size.}
    \label{fig88}
\end{figure}

As an example the time evolutions of $\langle D(t) \rangle$ for $D(0) =~\pm 7, N=26$ and
two different interaction strengths, namely $\kappa = 3, 4.5$, are displayed in Fig. \ref{evo} 
First of all one should note that the non resonance condition and the dilute eigenstate occupation principle obviously
apply since in all instances
$\langle D(t) \rangle$ converges against constant values with time. However, 
while the energy difference  $\langle D(t) \rangle$ clearly vanishes quickly for $\kappa = 3.0$ (as required by the
validity of the ETH) this is not the case at
 $\kappa = 4.5$. There it appears that a fixed portion of the initial value persists. Hence, this system indeed
partially preserves an uneven distribution of 
 local energy in the long time limit, which may be viewed to be at odds with the principle of heat
conduction. This finding motivates the introduction of
the ``ISI quantifying'' parameter $\Lambda$,
which we define as the long time equilibrium value divided by the initial expectation value. Since the
preserved portion seems to be more or less independent of actual $D_0$, we calculated only dynamics featuring $D_0 =
N_L - 2$. Thus, the definition of the "ISI quantifying" parameter reads
\begin{equation}
\Lambda=\frac{{ \langle \phi_\text{MOD}| D(t) | \phi_\text{MOD} \rangle }}{\langle \phi_\text{MOD}| D(0) |
\phi_\text{MOD} \rangle}~~\text{with}~~t \gg \tau~,
\end{equation}
where $\tau$ is the relaxation time. Obviously $\Lambda=0$ indicates ISI whereas larger $\Lambda$ indicate a violation
of ISI. {Furthermore, note that $\Lambda$, like $v$, is insensible with respect to scaling of
observables
with system size and hence applies to any kind of observable.} We
computed  $\Lambda$ for three different 
interaction strengths $\kappa$ for increasing system sizes. The result is displayed in Fig. \ref{fig88} 
Clearly $\Lambda$ vanishes in the limit of large system for $\kappa=3.7$, thus ISI relaxation holds at this
interaction strength. At and above  $\kappa=4$, 
$\Lambda$ appears to converge against non-zero values, hence no ISI relaxation exists at these interaction strengths,
not even
for $N \rightarrow \infty$. This result should 
be discussed in relation to the results of Sec. \ref{ethinvest}. Obviously the transition form ISI to non-ISI
with increasing interaction strength happens at 
the same point at which $v$ starts to converge against non-zero values. Thus the behavior of $v$ predicts ISI
rather precisely. $\Sigma$, on the other hand, 
 starts to converge against non-zero values at   $\kappa>3.5$, i.e., in a regime in which ISI is still
present. Thus, in
this sense, $v$ appears to be a more 
 reliable predictor of ISI than $\Sigma$. This is one of the main results of this paper.

\section{Integrability investigations}
\label{integ}
 So far, we focused on the issue of ISI equilibration of a specific observable and found 
 the existence of two regimes, ISI and non-ISI, depending on the interaction strength. Next we address the existence of
an integrable
 regime in the model at hand and study its relevance for the emergence of ISI for the MOD states. Again, similar to the
discussion of ETH and ISI, 
 a range of papers more or less explicitly states that non-integrability is imperative for ISI \cite{Srednicki}, whereas
other works analyze ISI without even 
 mentioning integrability. Also different features  of ``statistical relaxation'' ( other than ISI) are
addressed; examples exist in which the occurrence of statistical 
 relaxation does not depend on integrability \cite{santos}. However, this type of investigation generally suffers from a
conceptual shortcoming: there is
no unambiguous definition for integrability in quantum mechanics. Contrary to classical mechanics notions like phase
space, Lyapunov exponents and
ergodicity are not  well defined. In the context of lattice particle systems, quantum integrability is sometimes
associated with  being accessible by the
Bethe ansatz, i.e., the 1D quantum Heisenberg-chain with nearest neighbor interaction \cite{bethe}, the 1-D
$\delta$-function interacting Bose \cite{lieb} and the Fermi \cite{gaudin} gases are considered as integrable.
{According
to this later
definition the model considered in this paper is integrable.}  Nevertheless,
the break down of the ETH for large interaction strength may be viewed as being due to the proximity of the integrable
(according to any standard definition) 
limit of non-interacting spin-dimers. In order to address and quantify this possible integrability, we resort to the
well known approach  which is  based on 
the Nearest Neighbor  Level Spacing Distribution (NNSD) denoted by 
$P(\Delta\epsilon)$  \cite{wigner,dyson}. The distinction between integrability  and chaos is as follows: if the most
frequent energy spacing
is approaching zero $\Delta\epsilon=0$ and the shape of the  NNSD mimics a Poisson distribution, the system is
considered to be integrable. Whereas, if the most frequent energy spacing takes some finite value (level repulsion) with
an NNSD shape of Wigner-Dyson type, the system is considered to be non-integrable. This
classification of quantum systems using level statistics has been derived in the context of quantum models, whose
corresponding classical counterparts are chaotic \cite{gutzwiller} and has been adopted even for quantum systems which
do
not have classical counterpart,.e.g., spin systems. \\

\begin{figure}[h!]
   \centering
    \includegraphics[width=0.42\textwidth]{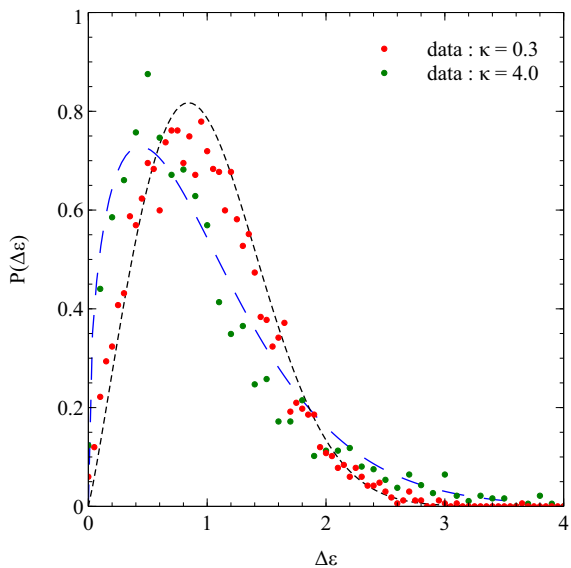}
    \caption{NNSD for $\kappa = 0.3$ and $\kappa = 4.0$ both for $N = 14$ where symbols display computed data and dashed
lines corresponding Brody distributions. In case of $\kappa = 0.3$ the Brody parameter reads $\omega \approx 1.1$
(Wigner-Dyson
type) and
for $\kappa = 4.0$ the Brody parameter reads $\omega = 0.4$ (Poisson
type).}
    \label{fig5}
\end{figure}

\begin{figure}[h!]
   \centering
    \includegraphics[width=0.5\textwidth]{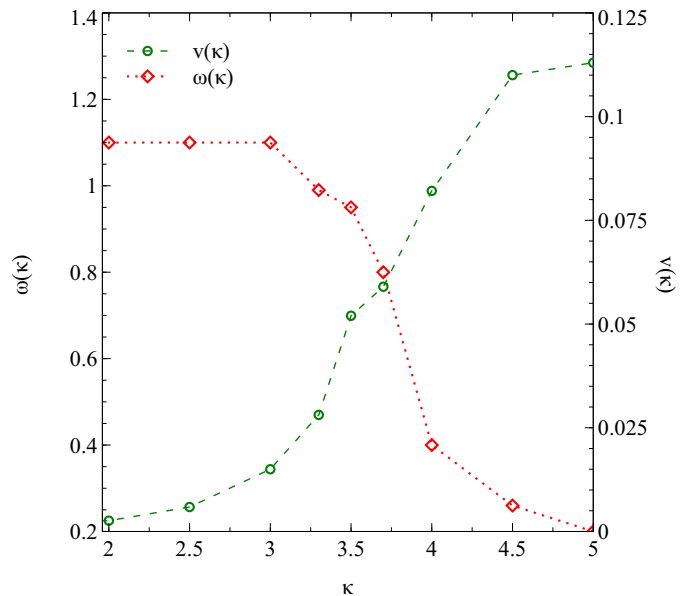}
    \caption{Comparison of Brody parameter $\omega$ and alternative ETH parameter $v$. The transition from integrable
to non-integrable systems indicated by $\omega$ obviously coincides with the transition from the ISI to non-ISI
regime, respectively. This links the equilibration dynamics to quantum chaos; see text for details. 
Note that that $\omega$ is gained from systems of $N=14$ whereas $v$ is gained from systems of size $N=20$.}
    \label{fig10}
\end{figure}

For most finite systems of condensed matter type the NNSD turns out not to correspond exactly neither to  Poisson nor
Wigner-Dyson like distributions. To deal with 
intermediate statistics, Brody
proposed in Ref. \cite{brody} to compare each real NNSD to a one
parameter $\omega$-family of analytically given  NNSD's, where $\omega=0$ corresponds to pure Poisson and $\omega=1$
to
pure Wigner statistics. Matching
a real NNSD to a pertinent Brody-NNSD thus yields a specific $\omega$ that may be used to quantify the closeness to
either Poisson or Wigner, respectively. 
Following this scheme, we computed (by means of exact diagonalization) and normalized an NNSD for various
interaction strengths and
$N_R=6$ in a narrow energy interval around $E=0$. The results, for  $\kappa = 0.3, 4.0$, together with the matching
Brody NNSD's are displayed in Fig. \ref{fig5}.
Obviously the agreement is rather good. This justifies the usage of the above described method to quantify the ``degree
of integrability''
by means of the parameter $\omega$. Finally, Fig. \ref{fig10} displays $\omega$ as a function of $\kappa$ together with
the ETH parameter $v$.
First one should note that for large interaction strengths the NNSD is much closer to Poisson than to Wigner. This
indicates that this model class may indeed
become integrable again for stronger interactions, say $\kappa \approx 3.7$, which are nonetheless far away from the
integrable dimer limit at $\kappa \gg 1$.
To repeat, this integrability is not induced by the possibility of applying a Bethe ansatz. We are furthermore unable to
judge whether this integrability 
may be explained by stretching the quantum KAM theorem \cite{Kang} all the way down from  $\kappa \rightarrow \infty $
to $\kappa \approx 3.7$.
However, it is striking that the transition from ISI to non-ISI (indicated by increasing $v$) happens at the very same
point at which the NNSD changes from 
 Wigner to Poisson (indicated by decreasing $\omega$). While this finding is just based on numerics in suggests that
``chaoticity'' in the sense of a large Brody 
 parameter may indeed be a sufficient criterion for the ISI relaxation of few-body observables for systems starting in
MOD states. Investigations on different spin 
 systems that point in a similar direction \cite{steinigeweg} also exist. 

{
\section{towards the  physics behind the numerical findings } \label{intpret}
While the numerical results clearly indicate a breakdown of ``chaoticity'' 
as well as of full relaxation of differences of local energies at strong couplings, the physical reason for this behavior is yet unclear. 
In the following we shortly speculate about such physical reasons, thus arriving at at some suggestions for further research.\\
All our findings refer to exchange of local energy between two asymmetric legs of a ladder. It may, however, be elucidating to consider the exchange of 
local energy between the part of the system that is a regular ladder (upper part in Fig. \ref{fig1}) and the part that consists of the  ``elongation'' of one leg
that really is just a chain (lower part in Fig. \ref{fig1}). If energy exchange between these two parts (ladder and chain) is suppressed, this will result 
in a suppression of energy exchange between the original  asymmetric legs, since the two energy-differences surely have an overlap  in the sense of a 
Mazur inequality \cite{mazur}. Thus taking the ``ladder and chain'' point of view, two features are evident: In the limit of  large $\kappa$  the mean 
level spacing in the ladder will eventually become significantly larger than the  mean level spacing in the chain.
Furthermore, the chain is integrable in the sense 
of a Bethe-Ansatz, whereas the ladder is not. Regarding the increasing level spacing in the chain, it may be the case that transitions that amount to an exchange of
local energy become more and more off-resonant. If the coupling strength (here: between ladder and chain, i.e., not $\kappa$) remains constant but the coupling 
becomes rather off-resonant, the relaxation of the respective observable may eventually not only be slowed down but inhibited completely 
\cite{occurrenceofstatisticalrelaxation}.  
Concerning the integrability, the eigenstates of the chain may in principle be described by a respective set of rapidities, whereas this is not applicable to 
the eigenstates of the ladder. This structural difference between the eigenstates may cause  the eigenstates of the chain to scatter strongly at the interface, thus 
preventing them from penetrating deeply into the ladder. This  could also lead to an effective suppression of the transitions that facilitate an exchange of
local energy. \\
Deciding which of the two above schemes is (if at all) dominantly responsible for the inhibition of the exchange of local energy is beyond the scope of 
the present paper. Future research, however, could focus on the ``ladder and chain partition'' and take more variables, other than just local energy 
differences into account.}

\section{Summary and Conclusion} \label{summary}
The paper at hand aims at clarifying the interrelations between the eigenstate thermalization hypothesis (ETH), initial
state independent relaxation and 
chaos in  quantum systems. The investigations are of primarily numerical character and focus on a class of
{(asymmetric)} ladder type
spin systems with variable interaction 
strength between the legs. {We investigated the energy difference between the legs of the spin system
where we found that two ETH quantifying parameters, a standard one and an alternative, recently
suggested one, indicate violation 
of the ETH, even  in the limit of large systems above certain ``threshold'' interaction strengths.} However, the thresholds
differ for the standard and the alternative 
ETH parameter.\\
Furthermore, the relaxation behavior of the energy difference between the legs of the spin system is
analyzed for a specific class of 
initials states. It is found that those energy differences no longer equilibrate to zero above a certain interaction
strength. This interaction strength 
precisely coincides with the ETH violation threshold of the alternative ETH parameter but not with the threshold of the
standard one.\\
Finally the level spacing 
statistics are considered. It turns out that they shift from Poisson-type to Wigner-Dyson type with increasing
interaction
strengths, again rather distinctly  at 
the threshold of the alternative ETH parameter. To conclude: numerical evidence suggests that the alternative ETH
parameter reliably signals the relaxation 
of an observable towards a common value that is independent of the initial, possibly largely off-equilibrium value of
the respective observable. Moreover there 
appears to be a strong correlation of this alternative parameter with either chaotic or integrable type of level statistics.


%

\end{document}